\documentclass[pra,aps,twocolumn,showpacs,groupauthors]{revtex4}

\usepackage[english]{babel}
\usepackage[ansinew]{inputenc}
\usepackage{amssymb}                                                           
\usepackage{amsmath}                                                           
\usepackage{graphicx}                                                          
\usepackage{bm}                                                                

\begin{document}


\title{Quantum-limited position measurements of a dark matter-wave soliton}


\author{Antonio Negretti$^1$}\email[E-mail: ]{negretti@phys.au.dk}
\author{Carsten Henkel$^2$}\author{Klaus M\o lmer$^1$}
\affiliation{1. Lundbeck Foundation Theoretical Center for Quantum System
Research,\\
Department of Physics and Astronomy, University of Aarhus,
DK-8000 Aarhus C, Denmark\\
2. Institut f\"ur Physik, Universit\"at Potsdam, Am Neuen Palais 10, D-14476 Potsdam, Germany}
\date{27 March 2008}


\begin{abstract}
We show that the position of a dark matter-wave soliton can be determined with a precision that scales
with the atomic density as $n^{-3/4}$. This surpasses the standard shot-noise detection limit for
independent particles, without use of squeezing and entanglement, and it suggests that interactions
among particles may present new advantages in high-precision metrology. 
We also take into account quantum density fluctuations due to phonon and Goldstone modes and we show that they, 
somewhat unexpectedly, actually improve the resolution. This happens because the fluctuations depend on the 
soliton position and make a larger amount of information available.
\end{abstract}


\pacs{03.75.Nt; 06.20.Dk; 37.25.+k}
\maketitle


\section{Introduction}

Degenerate quantum gases of atoms and molecules can be trapped in space and their interactions can be
controlled by laser and magnetic fields. The spatial quantum state in these gases can be probed in
ways unparalleled in other many-body systems. A number of experiments have thus been devoted to
exploring their coherence properties by the observation of interference patterns \cite{Andrews1997} and
topological excitations such as solitons and vortices \cite{Burger1999,Madison2000,Dutton2001}. For
application of ultra-cold gases in high-precision metrology, it is a central 
question how to
extract the maximum information from the accessible measurements.  We
address the case of spatial density measurements on a Bose-Einstein
condensate (BEC) with a dark soliton dip, which is a minimum in
the atomic density profile. 
Dark solitons have been produced by slowing down a
light pulse using electromagnetically induced transparency (EIT)
\cite{Dutton2001}, and the precision probing of the position of such
solitons may be used to measure detailed properties of EIT.

It has also been proposed, with a similar setting as in Ref.~\cite{Jo2007}, 
to use a time-dependent atomic trap
guide as a matter-wave interferometer where, after splitting and recombining of a BEC, an oscillating dark
soliton is formed~\cite{Negretti2004}. A high-precision determination 
of the soliton position is crucial here since
the amplitude of the soliton oscillation depends on the
interferometric phase.
The signal-to-noise ratio is in many interferometric experiments given by the standard shot
noise, leading to a phase resolution that scales with $1/\sqrt{N}$, where $N$ is the number of
detection events. The normal interference fringes in a matter-wave interferometer are located in
accordance with the interferometer phase and yield the same dependence on the number density. Special
quantum features such as entanglement and squeezing imply that this resolution limit is not a
fundamental one \cite{Giovannetti2004}. In principle, the Heisenberg uncertainty relation provides the
optimum sensitivity with a scaling as $1/N$~\cite{Bouyer1997}, and for particular entangled states the
interferometric phase error scales as $1/N^{3/4}$\cite{Pezze2007}. Our work will show that the
position of a dark soliton can be determined with a precision scaling 
with the same power law.
This result, however, does not rely on squeezing or entanglement. It can be qualitatively understood
from counting shot noise, the slope of the mean density versus position, and the $1/\sqrt{n}$
dependence of the soliton width on atomic density due to interactions.
Our analysis determines the Fisher information available in the spatial recording of the atomic
density and the resulting Cr\'amer-Rao lower bound (CRB) \cite{Refregier2004} on the determination
of the position of the soliton dip.

\begin{figure}[t]
\begin{center}
\includegraphics[width=8.6cm,height=12.90cm]{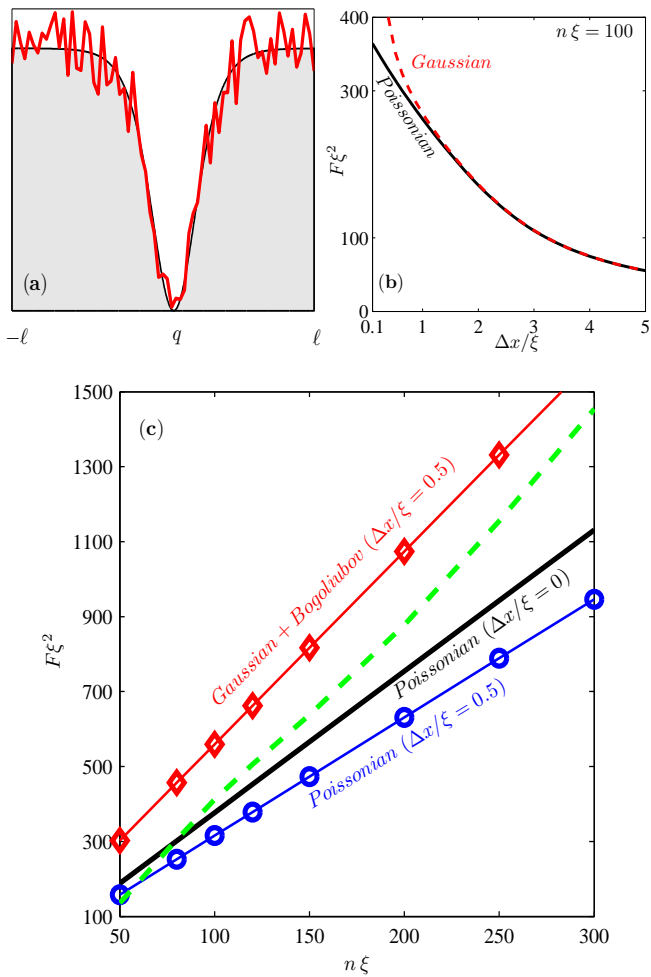}
\end{center}
\caption{(Color online). (a) Dark soliton density distribution (shaded region) and
typical measurement outcome with number fluctuations (red thick curve). (b) Rescaled Fisher
information $F \xi^2$ versus pixel size for a mean-field solution with Poisson counting
statistics (solid curve) and with a Gaussian continuous approximation (dashed curve).
The optimum sensitivity for locating the soliton position is $\Delta 
q = F^{-1/2}$; $\xi \propto n^{-1/2}$ is the healing length.
(c) Rescaled Fisher information versus atomic density:
Poissonian distribution (circles), Bogoliubov and Gaussian approximations (diamonds).
Continuous thick (black) line
represents the behaviour of Eq. (\ref{Eqn:FisherPois}), whereas the dashed (green) line shows
the information extracted from the signal-to-noise ratio for a simple gain function $g(x,n)$
(see text).
} \label{fig:SolitonSketch}
\end{figure}


\section{Spatial image of a dark soliton}

The experimental situation we have in mind is sketched in Fig. \ref{fig:SolitonSketch}(a). A quasi
one-dimensional (1D) BEC is imaged by counting the number of atoms present in sub-intervals,
referred to as pixels in the following, spanning the extent of the system. For simplicity we imagine
the probing to occur with unit efficiency, and we note that the recorded data will fluctuate around
the quantum-mechanical expectation value of the number of atoms present in each pixel.

A dark soliton is a stationary excited state solution of the
Gross-Pitaevskii equation (GPE)~\cite{Pitaevskii2003},
\begin{eqnarray}
\label{Eqn:GPE}
0 = \mathcal{H}_{\rm GP}\,\Phi
\equiv -\frac{\hbar^2}{2\,m}\frac{\partial^2 \Phi}{\partial x^2}
+ g\,|\Phi|^2\,\Phi - \mu\,\Phi,
\end{eqnarray}
where $\Phi(x)$ is the order parameter, $m$ is the atomic mass, $g$ is the effective 1D
interaction strength, and $\mu$ is the chemical potential.
For finite systems $\Phi$ is normalized to the number of particles in the gas,
and is proportional to the single-particle wave function, self-consistently occupied by all the
atoms in a Hartree-Fock product state. When $g>0$ (repulsive interactions) the
dark soliton in the homogeneous case is given by \cite{Pitaevskii2003}
\begin{eqnarray}
\label{Eqn:Soliton}
\Phi(x; q) = \sqrt{n}\tanh\left[\kappa (x - q)\right],
\end{eqnarray}
where $\kappa = 1/(\sqrt{2}\,\xi)$ is the reciprocal width of the
soliton profile, $n$ is the linear density of the condensate, and $\xi = \hbar/\sqrt{2\,m\,g\,n}$ is the
healing length. The position of the soliton is described by the parameter $q$.

When dealing with not truly 1D Bose gases, even in the dimensionality cross-over \cite{Theocharis2007},
care has to be taken. Exact 1D solitons are stable because of the integrability of the 1D GPE, 
while solitons in 3D elongated condensates are thermodynamically unstable
because of scattering with thermal excitations \cite{Muryshev2002}. We consider our 1D model to be
a good description for the soliton dip uncertainty, especially when a measurement is performed shortly
after its preparation \cite{Dziarmaga2004}. We note also that part of our analysis can be readily 
generalized to the full 3D description and also to effective 1D descriptions 
with a suitably modified 
GPE~\cite{Salasnich2002,Mateo2008}.

The number $n_s$ of atoms in pixel $s$ will on average be equal to
the integral of the density
$\rho(x; q) = |\Phi(x; q)|^2$ over the pixel area $\Delta x$:
$\bar{n}_s = \int_{x_s}^{x_{s}+\Delta x}{\rm d}x\,\rho(x; q)$.
Due to fluctuations around the quantum-mechanical expectation
value, the counting data will be noisy, and we need a theory that
provides the best estimator for $q$.
The Cram\'er-Rao bound provides a lower limit on the minimum uncertainty of a parameter $X$,
determined from measured values of a quantity $\eta$, whose conditional probability distribution is
given by $p(\eta|X)$.  The variance of an unbiased estimator of $X$ is limited from below by
$\min[{\rm Var}(X)] = 1/F(X)$, where
\begin{eqnarray}
\label{Eqn:FisherInfo} F(X) = - \int\!{\rm d}\eta\,p(\eta|X)\, \frac{\partial^2\ln p(\eta|X)
}{\partial X^2}
\end{eqnarray}
is the Fisher information \cite{Refregier2004}. Let us assume that $q$
is known (by a coarse scale estimate) to within
an interval for which a first-order series expansion
$\sqrt{\rho(x; q)} \simeq \sqrt{\rho(x; 0)} + q\, f(x)$ is precise enough.
A mean-field condensate, whether in a total number eigenstate or a coherent state,
displays Poissonian counting statistics in each pixel, when registered in a large number of them.
An estimator that reaches the CRB is found by weighting the
measured data $n_s$ with a weighting function $g(x) \propto f(x)/\rho(x;0)$~\cite{Delaubert2008}.
We thus have a procedure by which we can estimate the soliton position optimally.

To identify the CRB we need to evaluate the Fisher information (\ref{Eqn:FisherInfo}),
where $\eta$ is the entire set of (discrete) variables $\{n_s\}$
with a probability
distribution that depends on the soliton position $q$. The Fisher information can be determined
analytically, as done in Ref.\cite{Delaubert2008} for a coherent state of light
populating a mode with a given spatial density.
For our dark soliton (\ref{Eqn:Soliton}) we get
\begin{align}
\label{Eqn:FisherPois}
F = 4 \int\!{\rm d}x \left[\frac{\partial \sqrt{\rho(x; q)}}{\partial q}\right]^2
= \frac{ 16 \sqrt{m g} }{3 \,\hbar} n^{3/2}.
\end{align}
This dependence on the density is shown in Fig. \ref{fig:SolitonSketch}(c) (continuous thick line).
The main result is that the smallest value of $q$
that can be distinguished scales as $n^{-3/4}$ with the atomic density. This is a faster convergence
with $n$ than the usual shot-noise scaling ($\sim n^{-1/2}$). We emphasize that this enhancement
does not require any squeezed or otherwise entangled multiatom state; it follows from a simple
(coherent state) macroscopic wave function.
Note also that we obtained Eq.(\ref{Eqn:FisherPois}) by assuming no correlations between
different pixels and taking the limit of infinitely small pixel areas.
Fig. \ref{fig:SolitonSketch}(b) (thick solid line) shows the dependence of the rescaled Fisher information
$ F\,\xi^2$ on the pixel size (placing a soliton dip at the border between two pixels).

The dark soliton owes its existence to interactions, but interactions also cause deviations of the
exact quantum many-body state from the simple mean-field solution. These
deviations cause quantum
depletion of the condensate and fluctuations of the soliton dip position, which cannot
simply be set to a classical value. We analyze these effects by Bogoliubov 
theory, and show that, contrary to what may be expected, quantum 
fluctuations actually increase the optimum sensitivity of the soliton position 
measurement.

First, we note that the case of Poissonian counting statistics is tractable because the integral (sum)
in Eq.(\ref{Eqn:FisherInfo}) over all possible measurement outcomes can be performed analytically.
Poisson statistics is fully parametrized by the mean values of the observables, and hence it does not
offer the possibility of investigating the role of extra noise or correlations in the gas. We
therefore establish a formalism in which it is possible to deviate from Poisson statistics and in
which a good approximation to the Fisher information (\ref{Eqn:FisherInfo}) can be computed
efficiently. As a natural choice we consider a multi-variate Gaussian approximation for the
distribution function of the atomic density, namely
\begin{eqnarray}
\label{Eqn:Gaussian} p({\mathbf n}|q) =
\frac{(2\,\pi)^{-M_{\rm px}/2}}{\sqrt{{\rm det}({\mathbf C})}}\,\exp\left[ -
\frac{1}{2}\, ({\mathbf n} - \bar{{\mathbf n}})^{{\rm T}}\,{\mathbf
C}^{-1}\, ({\mathbf n} - \bar{{\mathbf n}})\right].
\end{eqnarray}
Here the column vectors ${\mathbf n}$ and $\bar{{\mathbf n}}$ collect the
detected stochastic atom number variables and their expectation values,
and $M_{\rm px}$ is the total number of pixels.
The correlations between atom numbers on the pixels are described by
the covariance matrix ${\mathbf C}$.

Inserting (\ref{Eqn:Gaussian}) in (\ref{Eqn:FisherInfo}) gives the Fisher information as
\begin{align}
\label{Eqn:generalF}
F &= \frac{1}{2} \left\{
\frac{ \partial^2 {\det}({\mathbf C}) / \partial q^2
      }{ {\det}({\mathbf C}) }
      -\left[
\frac{\partial\, {\det}({\mathbf C})/\partial q}{{\det}({\mathbf C})}
\right]^2 \right.
\\
&\phantom{=} + \left.
\sum_{s,j}\left[
\frac{\partial^2 (\mathbf{C}^{-1})_{s\,j}}{\partial q^2}\,C_{s\,j} +
2\,(\mathbf{C}^{-1})_{s\,j}\,\frac{\partial\bar{n}_s}{\partial q}\,
\frac{\partial\bar{n}_j}{\partial q}
\right]\right\}.
\nonumber
\end{align}
When Eq.(\ref{Eqn:generalF}) is evaluated for a continuous Gaussian distribution with the same
covariance as a discrete Poisson distribution,
$C_{s\,j} ={\rm Var}(n_s)\,\delta_{s\,j} = \bar{n}_s\,\delta_{s\,j}$,
we get the Fisher information
\begin{eqnarray}
\label{Eqn:FisherGauss} F = \sum_s \frac{1}{\bar{n}_s}\,\left(
\frac{\partial \bar{n}_s}{\partial q}\right)^2 + \frac{1}{2}\,\sum_s
\frac{1}{\bar{n}_s^2}\,\left( \frac{\partial \bar{n}_s}{\partial
q}\right)^2.
\end{eqnarray}
The first sum is equivalent to the Poisson result, while the second sum is a consequence of our Gaussian
approximation. In the limit $\bar{n}_s \gg 1$, where the distributions are similar, this latter term
becomes a negligible correction. For infinitely small pixels, however, the Gaussian and
Poissonian distributions differ and the second term in Eq.(\ref{Eqn:FisherGauss}) diverges
as shown in Fig. \ref{fig:SolitonSketch}(b).
This divergence is linked to the nonphysical, noninteger, and negative values of $n$ allowed by  the Gaussian
distribution. In the analysis of real experiments, however, this is not relevant, since detectors will
typically not be chosen so small that only one or zero atoms are recorded in some pixels.


\section{Bogoliubov--de Gennes theory}

In order to determine the quantum fluctuations of the atomic density,
we have diagonalized the Bogoliubov--de Gennes operator
\begin{eqnarray}
\label{Eqn:Lgp}
\mathcal{L}_{\rm BdG} = \left(
\begin{array}{cc}
\mathcal{H}_{\rm GP} + g\, |\Phi|^2 & g\, \Phi^2\\
-g\, \Phi^{*2} & -\mathcal{H}_{\rm GP} - g\, |\Phi|^2
\end{array}
\right).
\end{eqnarray}
Here, the wave function $\Phi=\Phi(x; q)$ is the stationary mean-field solution given by
Eq.(\ref{Eqn:Soliton}).
Following the approach in Refs.\cite{Busch2000a,Dziarmaga2004}, we find
that the phonon mode functions $\left(u_k,v_k\right)\equiv\left(W^+_k,W^-_k \right)$
of $\mathcal{L}_{\rm BdG}$ in a box with periodic boundary conditions can be written as
\begin{align}
\begin{split}
W^{\pm}_{k}(x; q) & = \frac{ M_{k} }{ \kappa }
\,e^{i\,k\,x}\,\left\{
k\,{\rm sech}^2[\kappa\,(x-q)] + {} \right.\\
&\phantom{=}
\left. \beta_{k}^{\pm} \left(
k / 2 + i\,\kappa\,\tanh[\kappa\,(x-q)] \right) \right\},
\end{split}
\label{eq:phonon-modes}
\end{align}
where
$M_{k}$ is a normalisation constant
\cite{Negretti2007},
$\beta_{k}^{\pm}=\left(k/\kappa\right)^2\pm\,2\,\epsilon_{k}/(g\,n)$, and $\epsilon_{k}=\hbar \,c
\,|k|\,\sqrt{1 + k^2/(4\,\kappa^2)}$.

Before discussing the contribution of the phonon modes, we note that $\mathcal{L}_{\rm BdG}$ in
addition has two gapless modes.  They are Goldstone modes, originating from the breaking of the
translational and the global U(1) phase symmetry if one assumes a definite
value of the displacement $q$, and, e.g., a real order parameter $\Phi(x; q)$.
The zero modes together with their adjoint mode functions are
\cite{Dziarmaga2004,Negretti2007} ($N_{0}$ is the mean atom number
in $\Phi$)
\begin{eqnarray}
\label{eq:zero-modes}
u_{q}( x; q )
&=& - i\,\kappa\,\sqrt{n}\,{\rm
sech}^2[\kappa\,(x-q)]
\nonumber\\
u_{\theta}( x; q ) &=& \Phi( x; q )
\nonumber\\
u_{q}^{\rm ad}( x; q ) &=& \frac{-i}{4\,\sqrt{n}}
\nonumber\\
u_{\theta}^{\rm ad}( x; q ) &=& \frac{
 \Phi( x; q ) + i\, x\, u_{q}( x; q )
}{2\,(N_0 + n / \kappa)}
\label{eq:adjoint-mode},
\end{eqnarray}
with the associated functions $v_{\alpha}( x ) = - u^*_{\alpha}( x )$ and $v_{\alpha}^{\rm ad}( x ) = u^{{\rm
ad}*}_{\alpha}( x )$.
The quantization box is much larger than $\xi$, and we
neglect small boundary corrections.

We are now in position to compute the mean atomic density and the covariance matrix $\bf{C}$ in the
Bogoliubov approximation. The total matter-field operator is split into $\hat{\Psi}(x) = \Phi(x) +
\delta\hat{\Psi}(x)$, with $[\delta\hat{\Psi}(x),\delta\hat{\Psi}^{\dagger}(y)] = \delta(x-y)$,
and where $\delta\hat{\Psi}(x) = \sum_{\alpha=\theta,q}
\hat{P}_{\alpha} u_{\alpha}^{\rm ad}( x )
- i\,\hat{\theta}_{\alpha} u_{\alpha}( x ) + \sum_{k}
\hat{b}_{k} u_{k}( x ) + \hat{b}_{k}^\dag v_{k}^*( x )$ with $[ \hat{\theta}_{\alpha}, \hat{P}_{\beta}]
= i\,\delta_{\alpha\!\beta}$ as in  Ref.\cite{Dziarmaga2004}, and
$[\hat{b}_k,\hat{b}^{\dagger}_p]=\delta_{k\!p}$.

The Hamiltonian of the quantized field theory, expanded to second order in $\delta \hat\Psi$ and
$\delta \hat\Psi^\dag$, contains quadratic terms $\hat P^2_{\alpha} / 2 m_{\alpha}$ with effective
masses $m_{\alpha}$, and the usual phonon number operator contribution.
Therefore, in the lowest-energy state (and in thermally excited states) of the system, there are no
correlations between the phonon operators and the zero-mode operators.

The average density of atoms is given by
\begin{equation}
\langle\hat{\Psi}^{\dagger}(x)\hat{\Psi}(x)\rangle =
|\Phi( x )|^2 + \sum_{k} |v_{k}(x)|^2
+ \mathcal{Z}(x),
\label{eq:ave-density-BdG}
\end{equation}
where
\begin{eqnarray}
\mathcal{Z}(x) &=&  \sum_{\alpha=\theta,q} \left\{
|u_{\alpha}^{\rm ad}( x )|^2\,\langle\hat{P}_{\alpha}^2\rangle
+ |u_{\alpha}( x )|^2 \,\langle\hat{\theta}_{\alpha}^2\rangle\right.
\label{eq:Dq-gen}
\\
&&\left.
{} -u_{\alpha}^*( x )\,u_{\alpha}^{\rm ad}( x )
-2\,\Im[u_{\alpha}^*( x )\,u_{\alpha}^{\rm ad}( x )]\,\langle
\hat{P}_{\alpha}\,\hat{\theta}_{\alpha}\rangle\right\},
\nonumber
\end{eqnarray}
and the dependence on $q$ has not been displayed explicitly.
Given the modes (\ref{eq:zero-modes}), for real $\Phi$, the last term in (\ref{eq:Dq-gen}) vanishes, and the
zero-mode contribution to the density \emph{at the soliton position} becomes
\begin{align}
\begin{split}
\mathcal{Z}(q) &= \frac{n\,q^2\,\kappa^2}{4\,(N_0 + n / \kappa)^2}
    \,\langle \hat{P}_{\theta}^2 \rangle
    +
    \frac{ \langle \hat{P}_{q}^2 \rangle }{ 16\, n }
    +
    n\,\kappa^2 \,\langle \hat{\theta}_{q}^2 \rangle
    -
    \frac{\kappa}{4}.
\label{eq:Dq}
\end{split}
\end{align}
The operator $\hat P_{\theta}$ reflects the
fluctuation in the total condensed atom number. Poisson statistics thus implies $\langle
\hat{P}_{\theta}^2 \rangle\simeq N_0$. The soliton position is described
by the operator $\hat \theta_{q}$, and behaves like a free particle. We adopt here
the strategy, introduced in Ref.\cite{Dziarmaga2002} for a trapped BEC, 
choosing the quantum state for the $q$ mode to be the
unique Gaussian state in $\hat P_{q}$, $\hat \theta_{q}$ that
minimizes the average density
$\overline{n}( q ) = \langle\hat\Psi^\dag( q ) \hat\Psi( q )\rangle$.
This state has a finite spread $\langle \hat P_{q}^2\rangle$, so it evolves
slowly in time, and we consider measurements done quickly, before the spreading of the
soliton ultimately invalidates the Bogoliubov approach.

The density correlation function $\mathcal{C}(x,y; q)$ is found by a
straightforward expansion of field operator products to second
order in $\delta\hat\Psi$,
\begin{align}
&   \mathcal{C}(x,y; q) =
    \Phi( x ) \Phi( y ) \big\{
        \sum_{k} \left[u_{k}( x ) + v_{k}( x ) \right]
        \left[ u_{k}^*( y ) + v_{k}^*( y ) \right]
    \nonumber\\
&  \phantom{=} 
        {} + 4\left[
    \langle \hat P_{\theta}^2 \rangle
    u_{\theta}^{\rm ad}( x ) u_{\theta}^{{\rm ad}*}( y )
    +
    \langle \hat \theta_q^2 \rangle
        u_{q}( x ) u_{q}^{*}( y )
    \right]
    \big\}.
    \label{eq:result-density-correlations}
\end{align}
We finally get the elements of the covariance matrix $\bf{C}$ by integrating $\mathcal{C}(x,y; q)$
over finite-size pixels, $x\in [x_j,x_j+\Delta x]$ and $y\in [x_s,x_s+\Delta x]$.

The results of our numerical analysis are presented in Fig.\ref{fig:SolitonSketch}(c), which shows the
Fisher information for both the mean-field and the Bogoliubov descriptions of the condensate, where
the latter is computed within the Gaussian approximation.
All continuous lines show a linear dependence, which implies in our
dimensionless units that the soliton position $q$ can be probed
with a sensitivity that scales as $n^{-3/4}$. In
Fig.\ref{fig:SolitonSketch}(c), the dashed line
represents the
information that can be extracted from the signal-to-noise ratio for a gain function
$g(x,n) = \left\{1 - \tanh^2(\kappa x)[1+ \beta(n)]/2\right\}/\tanh(\kappa x)$, where
$\beta(n) = \tanh(0.014 n\,\xi - 0.84)$ represents a small correction to the optimal
weighting function for a mean-field condensate.

Why does the inclusion of noise in the Bogoliubov description provide an
even better resolution than the mean field? To understand this,
consider as in Ref.\cite{Delaubert2008}
the mean value of the detected signal 
$\bar{S}\simeq q\,\int\,{\rm d}x g(x)\partial_q\rho(x; 0)$
(for small $q$), and its variance
$\Delta S^2=\int\,{\rm d}x {\rm d}y g(x) g(y) \mathcal{C}(x,y;0)$.
The amount of information that can be extracted from this
estimation strategy is related to the signal-to-noise ratio
$\bar{S}^2/\Delta S^2$.
It can be shown from Eq.(\ref{eq:result-density-correlations}),
using the completeness relation for phonon and Goldstone modes,
that $\Delta S^2 = \Delta S_{\rm MF}^2 + \Delta S_{\rm ph}^2 - \Delta S_{\rm G}^2$ for any
gain funtion $g(x)$.
The Goldstone and phonon contributions, $\Delta S_{\rm G}^2$ and
$\Delta S_{\rm ph}^2$,
are of the same order in $n$, but $\Delta S_{\rm G}^2$ dominates
and we get smaller noise \cite{Negretti2007}.


\section{Conclusions}

In conclusion, we have used signal processing theory to show that the position of a dark soliton can
be measured with a precision that scales more favorably than the usual shot noise. Our study thus
illustrates that atomic interactions can enhance the performance of atom interferometers. Scaling
below the shot-noise limit is also reported for entangled or squeezed states, but such states are not
required in our scheme. Note that a similar potential for high-precision sensors may be achieved with
bright solitons \cite{Vaughan2007} and with vortex lattices, or with a ``bubble'' floating in a BEC and
filled with another atomic species \cite{Bhongale07}. The $n^{-3/4}$ scaling can be understood by
simple statistical arguments applied to the mean-field condensate. Our quantitative analysis by means
of the Cr\'amer-Rao bound shows, however, that
quantum fluctuations due to phonon
and Goldstone modes enhance the sensitivity because these modes 
provide density correlations that are sensitive to the soliton position
and make a larger amount of information available.


\section*{Acknowledgments}

The authors A.N. and K.M. acknowledge financial support from the European Union Integrated Project SCALA, 
and K.M. acknowledges support from the ONR MURI on quantum metrology with atomic systems. The author C.H. 
thanks the Deutsche Forschungsgemeinschaft for support (Grant No. He 2849/3). We thank U.~V. Poulsen for 
his comments on the manuscript.


\bibliographystyle{apsrev}
\bibliography{NegrettietalPaperNewBis}

\end{document}